# DESCRIBING EMERGENCY REMOTE TEACHING USING A LEARNING MANAGEMENT SYSTEM: A SOUTH AFRICAN COVID-19 STUDY OF RESILIENCE THROUGH ICT

Ammar Canani, CITANDA, University of Cape Town, South Africa, cnnamm001@myuct.ac.za

Lisa F. Seymour, CITANDA, University of Cape Town, South Africa, lisa.seymour@uct.ac.za

**Abstract:** In an effort to counter the spread of COVID-19 many schools were forced to shut down. Primary schools in South Africa were forced to shift to emergency remote teaching abruptly relying on using a Learning Management System (LMS) to aid their teaching. LMSs helped primary schools build resilience to cope with unexpected events. An opportunity rose to study the affordances and constraints faced when using a LMS for remote teaching, specifically for primary school learners ('Gen Z') – a largely ignored area of research. Through a case study of 6 schools, this research describes the affordances and constraints of the LMS supported teaching system in use in primary schools. Affordances related to schools, learners and teachers while constraints were classified from a financial, technological, school, learner and teacher perspective. Noteworthy affordances included using LMSs as notice boards and satisfying parents, the key stakeholders. In terms of constraints, the digital divide was a recurring theme while device and data costs were consistently a limitation. There were many cases of schools not realizing the full potential of LMS usage. This research should be useful for primary schools wanting to use a LMS for their teaching.

## 1. INTRODUCTION

COVID-19 was declared a pandemic leading South Africa to a state of national disaster causing schools to shut down more than once (Morais, 2020; Ncwane, 2020; Ntsabo, 2020; Spinelli & Pellino, 2020). Primary schools were forced to abruptly shift to emergency remote teaching (ERT) without preparedness (Ncwane, 2020) relying on learning management systems (LMS) to aid in teaching (Van Wyk, 2020).

A LMS is used to manage, create and deliver content, manage courses, create automated tests, track learner progress (Kraleva, Sabani, & Kralev, 2019) and facilitate communication between teachers and learners (Kasim & Khalid, 2016) while removing the need for physical presence as it available at any time and place using an internet connection (Walker, Lindner, Murphrey, & Dooley, 2016).

The traditional classroom setting does not meet the interest of primary learners, 'Gen Z' (born after 1996), increasing the need to have digital tools for learning (Karthikeyan, Rajasekaran, & Unyapho, 2019). With the emergence of new technologies, teaching methodologies have advanced adding pressure on LMSs to adapt (Kraleva, Sabani, Kralev, & Kostadinova, 2020; Sabariah, Santosa, & Ferdiana, 2019) such as enabling mobile learning (Kongsgården & Krumsvik, 2016). Primary school learners show positive attitudes towards using a LMS in their learning as mobile learning has improved learning and teaching (Karalar & Sidekli, 2017). A large gap exists in the literature for primary learners (Gen Z) that hasn't been researched extensively such as LMS and mobile learning (Haßler, Major, & Hennessy, 2016; Karalar & Sidekli, 2017; Kraleva et al., 2020; Sabariah et al., 2019). Little is known on how organizations use technologies during a crisis (Mirbabaie, Bunker, Stieglitz, Marx, & Ehnis, 2020; Sakurai & Chughtai, 2020). Most studies reviewed by the researchers focused on tertiary education.

This research aimed to answer the question: What are the affordances and constraints of using a LMS in South African primary schools for remote teaching during the COVID pandemic? This research did not use any theoretical framework and themes were inductively derived. The research can assist any educational institution looking to move from a traditional classroom environment to incorporating a LMS in teaching. Management or executive teams in schools will through learning from other's experiences, maximise affordances and reduce constraints with LMS usage. The next section reviews the literature, followed by the research method used, the findings and ending with the conclusion.

## 2. LITERATURE REVIEW

To understand the affordances and constraints of using a LMS, this literature review will touch on ERT that increased the reliance on using a LMS (Van Wyk, 2020), the background of LMSs and literature on affordances and constraints of LMSs.

### 2.1. Emergency Remote Teaching

Due to COVID-19, ERT was implemented providing an alternative way of delivering education during a crisis (Hodges, Moore, Lockee, Trust, & Bond, 2020). ERT aims to give quick access to learning material as opposed to a robust environment for learning (Pohan, 2020). ERT is a short-term solution to creatively find a way to continue teaching (Hodges et al., 2020). During 2020, South Africa launched the 'COVID-19 Learner Support' initiative providing COVID-19 TV and Radio stations (SABC, 2020) and resources on zero-rated platforms (Department of Basic Education, 2020).

### 2.2. Resilience

Resilience refers to the ability of systems to cope with unexpected events (Heeks & Ospina, 2019). Resilience is mostly known as a system property and a response to an event (Heeks & Ospina, 2019). A LMS supports teaching, and hence is part of the teaching system providing resilience to schools (Kraleva et al., 2019; Sarkar, Wingreen, & Ascroft, 2016). Therefore, a LMS provides resilience of an information system outcome system (RISOS) (Heeks & Ospina, 2019).

### 2.3. Background of LMS

The LMS originated at the University of Illinois in 1960 with Programmed Logic for Automated Teaching Operations (PLATO) and 'LMS' was coined as the management of the system (Chaubey & Bhattacharya, 2015). It began as Integrated Learning System (ILS) providing instructional content (Watson & Watson, 2007). In 1990, the first software based LMS was launched (Chaubey & Bhattacharya, 2015) and implementation grew in the early 2000s with the internet (Roy, 2017). Currently, a LMS, also called an online learning platform (Aldiab, Chowdhury, Kootsookos, Alam, & Alam, 2019), describes many systems working together providing online learning services.

### 2.4. Affordances of a LMS

An affordance refers to a completed action between a user and an IT artifact which emerges after an active exploration from a user in a certain context (Lanamäki, Thapa, & Stendal, 2017). Affordances do not refer to canonical features of an artifact but how they are 'seen as' by the user (Lanamäki et al., 2017).

The literature refers to four affordances from features provided by a LMS – accessibility, interactivity (Holmes & Prieto-Rodriguez, 2018), scalability and standardization (Ramírez-Correa, Rondan-Cataluña, Arenas-Gaitán, & Alfaro-Perez, 2017). Accessibility refers to being able to access content at any time and place using an active internet connection (Aldiab et al., 2019; Berking & Gallagher, 2016; Chaubey & Bhattacharya, 2015; Epping, 2010; Kasim & Khalid, 2016) while supporting different devices (Aldiab et al., 2019). Interactivity refers to having an attractive environment (Aldiab et al., 2019), user friendliness and ease of usage (Kasim & Khalid, 2016; Kraleva et al., 2019) providing on-demand content delivery facilitating interaction between teachers and learners (Berking & Gallagher, 2016). Scalability refers to integrating with external tools (Aldiab et al., 2019), exchanging data with other systems (Kraleva et al., 2019) and the reusing or combining of components of a LMS (Berking & Gallagher, 2016). Lastly, standardization refers to setting of tests, keeping learner records, tracking and reporting progress and identifying learners at risk (Berking & Gallagher, 2016; Kraleva et al., 2019).

Other affordances are referred to in the literature as benefits. For example, from the school's perspective teaching can continue online as a LMS allows remote learning eliminating the physical logistics (Mafuna & Wadesango, 2012; Ndobe, 2018). The LMS offers a central repository accessible from any place at any time (Berking & Gallagher, 2016; Holmes & Prieto-Rodriguez, 2018; Kraleva et al., 2019, 2020; Papadakis, Kalogiannakis, Sifaki, & Vidakis, 2017).

Mobile learning is enabled. Learners show positive attitudes towards using tablets and can use mobile phones although sometimes with poor accessibility (Karalar & Sidekli, 2017; Papadakis et al., 2017). Using tablets has shown to improve learners' academic performances and motor skills (Hubber et al., 2016). A LMS helps build resilience in learners and educational institutions as they are able to cope through unexpected events (Ayebi-Arthur, 2017; Heeks & Ospina, 2019; Wadi, Abdul Rahim, & Yusoff, 2020).

From the learner's perspective, it provides an appealing environment that is easy to use and is user friendly supporting gamification and catering to the demands of 'Gen Z' teaching techniques (Aldiab et al., 2019; Berking & Gallagher, 2016; Coates, James, & Baldwin, 2005; Kraleva et al., 2019, 2020). It allows to cater to learning methods of primary school learners, promoting learning subtly (Ting, 2019) to a point that learners "own the text" (Kongsgården & Krumsvik, 2016, p. 261). It empowers learners to work at their own pace and method of studying and equips them with technical skills (Kulshrestha & Kant, 2013).

Teachers get more time for planning, decreasing the administrative load. Teaching quality improves as teachers develop technical skills, reflect on their teaching methodologies and work collaboratively to overcome challenges while being able to experience content as a learner (Lonn & Teasley, 2009; Underwood, Cavendish, & Lawson, 2006; Unwin et al., 2010; Walker et al., 2016; Zheng et al., 2018). Teachers use various applications for teaching and learners become more creative with collaboration and group work, increasing active participation from learners (Kongsgården & Krumsvik, 2016). It shifts the focus of learning to the matter on hand leaving the topics out of scope to be handled by technology (Ben-Zvi, 2009).

### 2.5. Constraints of LMS Usage

A major constraint with LMS usage is cost. There are several costs that schools incur when using a LMS. Educational institutions opt for open-source LMSs, instead of licensed ones, to reduce maintenance and improvement costs (Anand & Eswaran, 2018; Kasim & Khalid, 2016). Yet free versions of LMSs have limited capabilities and additional costs are incurred to train staff members (Kraleva et al., 2019; Walker et al., 2016). Most people living in Sub-Saharan Africa live under $1.90 (R27) a day (United Nations, n.d.). 54% of South Africans do not own a computer, 62% have access to the internet using their phones of which 36% have never connected to the internet (Isbell, 2020). Therefore, the costs to provide devices and data to learners falls on the educational institutions such as the University of Cape Town or University of Witwatersrand that provided data and laptops to learners (Vermeulen, 2020).

Having access to the internet, good bandwidth and a personal computer to be able to access a LMS is a challenge for learners (Hillier, 2018; Mtebe, 2015; Unwin et al., 2010). Lack of ICT literacy creates barriers for using a LMS (Hillier, 2018). English not being the first language of learners affects the effectiveness of a LMS (Sackstein, Coleman, & Ndobe, 2019). Lack of computer laboratories in primary schools due to insufficient resources affects access (Mark & Emmanuel, 2019).

Due to a lack of time, teachers are not trained as it affects their time for planning thus hindering adoption of a LMS and a lack of technical support causes teachers to resist the new changes (Coleman & Mtshazi, 2017; Sackstein et al., 2019; Unwin et al., 2010; Walker et al., 2016; Zheng et al., 2018). Adoption usually comes from the top-down, increasing resistance from teachers (Mtebe, 2015). Even teachers that use a LMS do not use it to its full potential affecting return on investment (Bousbahi & Alrazgan, 2015).

LMSs are made in developed countries where children are exposed to technology at a young age (Sackstein et al., 2019). As of 2017, 22% of South Africans had access to ICT infrastructures (International Telecommunication Union, 2017). The gap has increased the digital divide in terms of social resources, cognitive resources and material resources (Lembani, Gunter, Breines, & Dalu, 2020). Due to lack of funding, most primary schools haven't been able to benefit from using a LMS (Chigona, Chigona, Kayongo, & Kausa, 2010).

To conclude, ERT is not fully understood (Hodges et al., 2020). With new technologies emerging, teaching methodologies have changed forcing LMSs to adapt (Kraleva et al., 2020). Primary school learners remain to be extensively researched with regards to using a LMS (Kraleva et al., 2019; Sabariah et al., 2019).

## 3.  METHOD

This section describes how the research was conducted while stating the assumptions and limitations. The purpose of this research was to describe the affordances and constraints of using a LMS for remote primary school teaching in times of crisis. The research question states 'what is' by describing and analysing the purpose of the research without explaining causality or making predictions (Gregor, 2006). An interpretive philosophy was adopted to understand the experience from the participant's perspective (Cussen & Cooney, 2017; Díaz Andrade, 2009). The assumption was that reality is a social construct that the researcher drives to reveal (Walsham, 1995). The philosophy helped give an insight about experiences from the perspective of the person living it (Díaz Andrade, 2009). The strategy of the research was a multiple-case study (Yin, 2009) where each school was a case. A case study was used as it provides insights not available from other strategies using multiple sources of data (Rowley, 2002). Prior to contacting any schools, permission was obtained from the university's ethics committee and the Western Cape Education Department. All participants consented to the research and were free to leave at any time. The research happened throughout 2020. The criterion used to select schools were:

- A primary school
- Based in South Africa
- Affected by COVID-19
- Used a LMS

Six primary schools based in Cape Town were chosen meeting the criterion. There were three public fee-paying schools, one public fee-paying school privately funded, one private fee-paying school and one semi-private fee-paying school. Each school had different infrastructure and accessibility levels to use a LMS. While most schools used Google Classroom there was a spread of the number of years a LMS had been used at the school. Table 1 lists schools selected with the ID being the case.

| ID | School Type | Infrastructure | Accessibility | LMS used | LMS Years |
|---|---|---|---|---|---|
| C1 | Public non-fee paying privately funded | Good | High | Google Classroom | 5 |

| C2 | Public fee-paying school | Average | Medium | Google Classroom | 6 |
| C3 | Private fee-paying school | Excellent | Very High | Google Classroom | 10 |
| C4 | Semi-private fee-paying school | Average | Medium | Google Classroom | 0 |
| C5 | Public fee-paying school | Average | Medium | Worksheet Cloud | 0 |
| C6 | Public fee-paying school | Average | Low | Google Classroom | 3 |

**Table 1. Schools Selected**

Judgement sampling was utilized to pick participants best fit to achieve the purpose of this research (Marshall, 1996) along with convenience sampling – selecting participants accessible for the research (Etikan, Musa, & Alkassim, 2016). The researcher was referred to schools by an expert working with LMSs in primary school. Teachers available and willing to be interviewed formed part of the sample. Participants were interviewed using online conferencing tools. Participants provided documentation to the researcher such as relevant policies. Semi-structured interviews with open-ended questions were done following an interview guide. The interview guide was developed using the research question in mind while being guided by the literature reviewed (Figure 1).

**Section 4: About Your LMS**

| YL01 | Which LMS does your school use? |
| YL02 | How long has your school been using an LMS? |
| YL03 | Does your school have support and maintenance for your LMS? |
| YL04 | Does your school offer training for the LMS? |
| YL05 | Are there any beneficial reports related to the LMS? May those documents be supplied after the interview? |
| YL06 | What is the perception of the students about the LMS? How has it changed due to COVID-19? |
| YL07 | What is the perception of the teachers about the LMS? How has it changed due to COVID-19? |
| YL08 | Do you gather any feedback from the users of the LMS? May those documents be supplied after the interview? |
| YL09 | How do you measure the effectiveness of your school's LMS? |

**Section 5: About Your LMS Usage**

| YU01 | What is your LMS used for by teachers and students? |
| YU02 | How do teachers and students access the LMS? |
| YU03 | Do you track the usage of the LMS by teachers and students? May those documents be supplied after the interview? |
| YU04 | How did the usage and reliance on the LMS changed during COVID-19? |

**Figure 1. Excerpt of the Interview Guide**

Interviews were recorded with consent from participants and later transcribed. Table 2 lists all participants interviewed. Participant IDs start with the case number and followed by the participant number from the case.

| ID | Role | Subjects | Teaching Years | LMS Years |
|---|---|---|---|---|
| C1P1 | Grade 5 teacher and administrator | English, Mathematics, History, Geography, Life Skills and Creative Arts | 5 | 3 |
| C2P1 | IT administrator and teacher | IT | 13 | 8 |
| C3P1 | Grade 4 teacher, school management team and basic level tech integration | Math and Science | 13 | 10 |
| C4P1 | Grade 7 teacher, department head | Social Sciences, Economics, Management Sciences, Life Skills and Natural Sciences | 9 | 0 |
| C5P1 | Grade 6 teacher | Mathematics and English | 13 | 0 |

| C6P1 | Grade 6 and 7 teacher | Afrikaans and Technology | 6 | 3 |

**Table 2. Participants Interviewed**

NVivo was utilized to code and analyse the data gathered. Transcripts of every interview were coded to find themes. Figure 2 shows initial coding. Themes were analytical outputs created by the intersection of the researcher's assumptions, analytical skills and resources and the data itself without using pre-existing codes (Braun & Clarke, 2019; Terry, Hayfield, Clarke, & Braun, 2017). Themes did not emerge from the data but were reflectively developed and generated (Braun & Clarke, 2019). To find themes, six phases of reflexive thematic analysis were used (Braun & Clarke, 2006):

1. Getting to know the data. The researcher became familiar with the data after transcribing and preparing it for analyses. This involved reading over the data more than once.

2. Generating initial codes. The researcher generated codes after each interview. The researcher then grouped common subjects together and reanalysed to generate more codes. Any point raised by a participant that touched on the research question was coded. Similar points were coded on the same code.

3. Finding themes. The researcher identified patterns emerging creating themes iteratively after each interview. Reviewing the existing codes, the researcher identified 'bigger-picture' themes that encompassed the codes.

4. Going over themes. Themes were reviewed to identify relevance to the theme and sub-themes were established. Themes were grouped into sub-groups as parent themes were generated from the existing themes.

5. Defining themes. Using gerund coding (Charmaz & Keller, 2016), themes were named to give explanation of the theme.

6. Generating a report. Using the data, the researcher wrote up the themes.

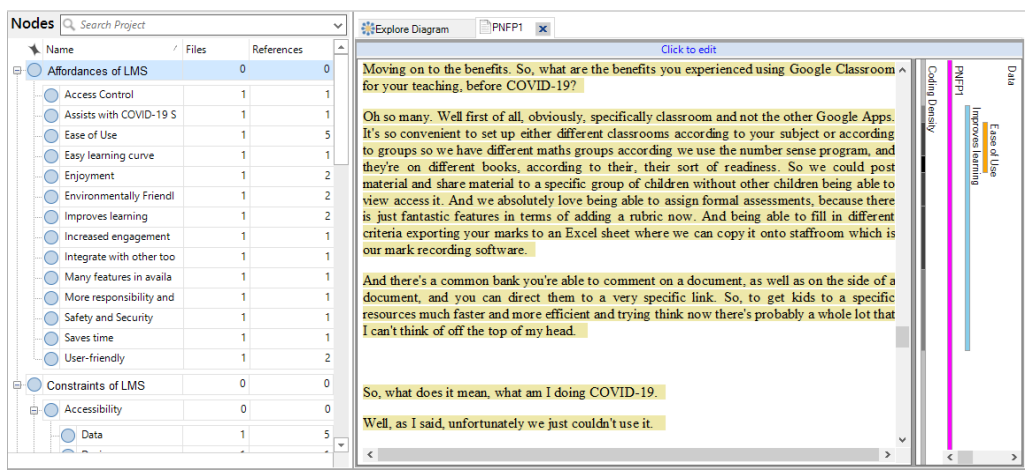

**Figure 2. Data Analysis on NVivo**

## 4. FINDINGS

Several themes were generated from the data collected, even though some themes overlapped with one another, it did not affect the classification of the finding. Contradicting themes were discussed at the first instance. The next section reports on the affordances and constraints of using a LMS. Affordances were classified as actions completed between a LMS and a user (Lanamäki et al., 2017). Constraints had a lot of overlap but the context specific to an actor was used as a guide for classification. Classifications were guided by the actor in the context of the theme emerging using the literature reviewed as a guide.

## 4.1. Affordances of a LMS

The affordances of a LMS have been classified based on the participant in the relevant LMS system or sub-system being the school (Table 3), the learner (Table 4) or the teacher (Table 5). These will now be discussed.

As classes were shut down, schools shifted to remote learning a key affordance of using a LMS. The literature refers to learners being able to access a LMS from any place at any time (Walker et al., 2016). A new affordance found was that a LMS enabled avoiding human and surface contact. For C2P2 their LMS became the only way of teaching – a new theme that was generated as literature shows that in primary schools a LMS is normally only used to supplement physical teaching (Berking & Gallagher, 2016). Going digital and paperless improved the storing and retrieving of documents – not initially found in the literature but a feature of information systems(Marsh, 2018). Some participants noted that they did not achieve the expected affordance of going paperless supporting literature's claim of this potential of using a LMS not always being realized (Bousbahi & Alrazgan, 2015). Improvement of teaching quality was in line with literature. Communication improvement through using a LMS, while noted in the literature was found to extend to using a LMS to send announcements to parents. Satisfying parent needs, an important consideration for primary schools (Teachwire, 2016), was identified as a new theme. Multiple school stakeholders could access content in their LMS with ease using different devices as noted in the literature. Flexible usage of a LMS is a new theme that was generated as literature states a LMS affords standardization but also presents challenges.

| Affordances of Schools | Data Evidence |
|---|---|
| Enabling avoidance of human and surface contact | [W]e are wanting to avoid cross contamination for example with marking books, and we wanting to limit the interaction of standing near learners, and, or, you know, touching their book their stationery etc. (C1P1). |
| Enabling remote teaching when physical classes are not possible | It's just, I mean, it (using a LMS) was the only way to teach (C2P2). |
| More efficient storage and retrieval | [W]e don't have to walk around with a planning file anymore. We don't have to search for videos or work notes anymore. I mean, it's right there (C4P4). |
| Helping improve teaching quality | [W]e can kind of see what works … what does not work and what we can fine tune (C4P4). |
| Improving Communication | [I]t (Google Classroom) became the only avenue of disseminating information so it became a vital part of our teaching and learning (C3P3). |
| Satisfying parent needs | Yes, it was actually very, very helpful, especially for the parent (C5P5). |
| Accessible from different devices | We (teachers) access via our notebooks, we picked up that some learners do it by their phones (C4P4). |
| Ease of use | It's (Google Classroom) user friendly (C3P3). |
| Providing a repository of information | [I]t's a storage place … your Google Classroom becomes your portfolio of what you do if you use it all the time (C3P3). |
| Allowing flexibility of usage | I think everyone in the school, every grade in the school uses [Google] Classroom in a different way… And I think the biggest challenge of that (using a LMS in different ways) was that not all of the grades in the school were using it the same way (C4P4). |

**Table 3. Data Evidence for Affordances of Schools**

From the learner's perspective, an easy learning curve reflects that a LMS caters for the demands of 'Gen Z' (Kraleva et al., 2020). Improvements in learning – an affordance of a LMS was consistent with the literature. Learners learning digital skills by using their LMS was noted in the literature.

Participants mentioned learners grasping concepts of responsibility and accountability – a new theme that was generated specific to primary school learners.

| Affordances of Learners | Data Evidence |
|---|---|
| Easy learning curve | That's it's normal to them (learners… (C2P2). |
| Improving learning | And that kids actually learn…(C2P2). |
| Learning digital skills | [W]e are moving towards more of a digital age, the skills, the digital skills that kids are learning (C3P3). |
| Learning responsibility and accountability | [Google Classroom] allow[s] the learner sometime after the work has been done, they can go and mark at home or they could go and revise them … they quite responsible now (C4P4 |

**Table 4. Data Evidence for Affordances of Learners**

From the teacher's perspective, the enabling of teaching and the saving of time by using a LMS were voiced by participants and noted in the literature. Motivating teachers to learn new technologies is a new theme that was generated since literature only refers to teachers learning digital skills. Having many features in their LMS afforded teachers to have everything in one place as noted in the literature.

| Affordances of Teachers | Data Evidence |
|---|---|
| Enabling teaching | Being able to still teach (C2P2). |
| Learning new technology | T]hey (teachers) will just dive straight in and we're (teachers) very happy to use it (Google Classroom) (C3P3). |
| Saving time | So, learners can check their work, which obviously saves us (teachers) time (C1P1). |
| Having many features available in one software application | The organization and having everything in one place (C2P2). |

**Table 5. Data Evidence for Affordances of Teachers**

## 4.2. Constraints of LMS Usage

Constraints have been classified as financial (Table 6), technological (Table 7), school (Table 8), learner (Table 9) and teacher (Table 10) constraints and are now described.

The financial constraints were costs incurred when using a LMS. These constraints applied to all users. The assumption was that all participants disclosed costs they were aware of – since all participants are teachers, the financial position of their schools is not their expertise. Several costs mentioned by participants such as devices, internet, licensing, support, and training were in line with the literature. C3P3 stated their school used the free version of Google Classroom as noted in the literature – enterprise versions have more capabilities and are usually not preferred by educational institutions. Increased printing costs was generated as a new theme contradicting a LMS's paperless affordance but remaining consistent with literature's stance on a LMS's potential not being fully realized.

| Financial Constraints | Data Evidence |
|---|---|
| Device Costs | [A]ll educators at home had the school's notebook… (C4P4). I use my cell phone or my laptop… (C5P5). |
| Internet Costs | So, my package, at home, is about 600 rand or just under, per month. And that's for high speed fibre (C1P1). We had a learner in my class, he says he spent about 300 Rand a week on data… (C4P4). |
| License Costs | I'm not sure what the license fee is to register for the number of staff we have using Google education (C1P1). The school has set that up for us (C4P4). For teaching well G Suite is a free education platform. So, that's the beauty of it, there are no costs involved (C3P3). |

| Printing Costs | [T]hey (learners) print the worksheets from [Google] Classroom and paste them in (C4P4). |
| Support Costs | Yes, we do use the company … who are our greater gurus who help us (C3P3). |
| Training Costs | [C1] has paid for Google training for all the teachers (C1P1). |

**Table 6. Data Evidence for Financial Constraints**

There were several constraints of the LMS technology experienced across all user groups. C1's LMS was unavailable for a period – a new theme that was generated to an extent of contradicting the availability affordance of an LMS but being consistent with constraints of any technology. Insufficient features in a LMS and poorly developed features remain consistent with literature as LMSs are built in developed countries. Educational institutions opt for free versions which come with limited capabilities too. C1P1 faced constraints in using other tools along with their LMS due to no integration capabilities – a new theme that was generated as literature states integration is an affordance of an LMS. Not being able to use a LMS offline limited students' access – a new theme that was generated requiring accessing a LMS without internet connection. A LMS requires an internet connection for accessibility, highlighting the effects of using LMS built in developed countries.

| **Technological Constraints** | **Data Evidence** |
| --- | --- |
| LMS unavailability | I think there was a period …where Google I've kind of shut down for a day or two. And we just then any lessons based on classroom and sharing material via classroom had to stop. And we had to come up with something else (C1P1). |
| Insufficient LMS features | I think it could be that deletions could be made available to the educators ahead of time (C5P5). |
| Lacking integration capabilities with other tools | And then with Google Slides you can't voice record over a presentation. So, then we'll have to use something like Adobe Spark, which doesn't integrate as nicely with Google Classroom as slides would, for instance (C1P1). |
| Poorly developed LMS features | Turn in function, which is silly (C2P2). |
| Requiring an active internet connection | So, the student doesn't need to have access (C5P5). |

**Table 7. Data Evidence for Technological Constraints**

There were many constraints that schools experienced with respect to LMS usage. The digital divide in South Africa was evident in the primary schools, responses from different schools varied, C1 stopped using their LMS while C3 was subsidizing internet costs for some students. C3P3's surprise at seeing their colleagues use WhatsApp groups instead of Google Classroom highlighted the digital divide – consistent with the literature. C2P2 mentioned lacking support from management, a new theme, that was generated although lacking support from management is a general constraint for any technology being introduced (Ismail, 2018). C4P4 noted timing constraints in setting up their LMS for ERT (a school without an existing LMS) – a new theme that was generated as the scope of this research did not include reviewing implementations of a LMS. Training users was time consuming, consistent with the literature. Some participants did not track usage of their LMS, an affordance of a LMS, due to having other priorities while others had their own strategy of tracking users – both reasons being consistent with literature as it takes time to train users and teachers to use a LMS to its full potential.

| **School Constraints** | **Data Evidence** |
| --- | --- |
| Digital Divide | During COVID-19 my biggest challenge was connecting to children who don't have Wi Fi access … they wouldn't do work on Google Classroom and I would have to WhatsApp the parents the work .... So that was a bit sad that not all the children could be join the classroom during the lockdown (C6P6). |
| Inadequate management support | They're (management) not super open to new tech ... Not much money for new tech either (C2P2). |
| Time needed to set up a LMS | I guess. I guess it's the initial setup … And then getting the parents to give us (teachers) a weekend to take off everything that we've learned and set it up to a point that makes sense for 124 learners plus four educators whose never used - that was a challenge …Yeah, so in the beginning, it was it was a tough sell (C4P4). |
| Time needed to train users to use LMS | S]o that was the main thing is showing everyone how to use it (C4P4). |
| Inadequate tracking of the usage by users | Oh, no, I don't think it is a priority right now (C1P1). For tracking – no. I am a teacher, so all of my time is spent on teaching. So no, that's something that the IT dude must do (C6P6). |

**Table 8. Data Evidence for Organizational Constraints**

In terms of constraints learners experienced, inadequate access to data and devices to use their LMS was referred to often, as noted in the literature. For this reason, C4P4 faced difficulty in introducing learners to their LMS. A new theme was generated, C4 rushed their introduction to their LMS leaving learners confused. Parents were unable to assist primary school students as they were lacking literacy skills – a new theme that was generated specific to primary school students while reiterating the effects of the digital divide. Even after contacting parents, C2P2 was struggling to ensure all students use their LMS highlighting a new theme contradicting the 'Gen-Z'-friendly image of a LMS in the literature.

| **Learner Constraints** | **Data Evidence** |
|---|---|
| Inadequate data access | And during the pandemic though our learners were in the category in South Africa who didn't have access to data (C1P1). |
| Lacking Device | [Learners] didn't all have access to a suitable smart device to use things like Google Classroom (C1P1). But a very few, a very small number of our students had access to computers and printing (C5P5). |
| Rushed introduction to an LMS | In the beginning, it was a bit, it wasn't easy. I think, the students, … they were a bit confused with that (features of Google Classroom) and they didn't know how to manage it (C4P4). |
| Inadequate parents' technological literacy | Every now and then we get a WhatsApp from parents saying they don't know where to find this or they don't know where to find that … but also parents and grandparents who can't even download the app (C4P4). |
| Unable to sign in to the LMS | [S]ome kids just, we came back to school after three months and they still hadn't joined the Google Classroom. After contacting parents and everything (C2P2). |

**Table 9. Data Evidence for Learner Constraints**

Teachers had poor connectivity which could have been resolved by upgrading their internet package – reflecting the digital divide (infrastructure) and financial constraints mentioned in literature. Consistent with the literature, teachers resisted using a LMS – some even went for training but never applied the concepts taught to them. Some teachers had inadequate technological skills – also an effect of the digital divide in South Africa.

| **Teacher Constraints** | **Data Evidence** |
|---|---|
| Inadequate Connectivity | We had Wi Fi; it wasn't as good as it could have been in a lot of homes (C3P3). I don't know if it is our internet facility or if it's because you're going through a third party but when we came back to school, signing up through classroom and then projecting on the board I think we found a bit of lag (C4P4). |
| Resisting using a LMS | I would say that the teachers were more reluctant to use it (C3P3). |
| Inadequate Technological Literacy | It's been a big learning curve (C4P4). It ranges from staff who are older, are not, you know, very comfortable with technology and who struggle with more basic functions, on, on a laptop. So, introducing something like Google has been a challenge for them (C1P1). |

**Table 10. Data Evidence for Teacher Constraints**

## 5.  DISCUSSION

The findings highlighted several themes consistent with literature. This section focuses on themes which were generated as affordances or constraints to the LMS enabled teaching system and were not evident in the literature. For affordances, a new affordance, specific to the COVID-19 context emerged as a LMS enabled avoidance of human and surface contact. In the same context, a LMS became the only way of teaching highlighting the role it played in increasing resilience in the teaching process in schools. Schools used a LMS to send announcements, extending the usage of a LMS for communication found in the literature. Flexibility in using a LMS, opposing standardization, brought its own challenges. Students learned accountability and gained responsibility indicating how a LMS may be used to increase resilience in students. LMSs motivated teachers to learn new technologies demonstrating the role LMS play in increasing resilience in teachers.

For constraints, printing costs were generated as a new theme but are captured as a LMS not being used to its potential. Constraints relating to support from management are general technological constraints. Learners faced constraints in using a LMS due to the rushed introduction to a LMS while their parents' technological literacy, a contribution specific to primary learners, constrained parental support to learners at home. Most constraints revolved around the existing digital divide in South Africa which was exacerbated by the pandemic. LMSs are advertised as 'Gen Z' friendly in contrast to findings of this research. This highlights the need for further research on LMS from the 'Gen Z' learner's perspective. The technological constraint of integration with a LMS was raised. This contradicts the affordance of a LMS and was due to varied brands being used for integration.

The themes that were generated in this research contributed to literature. From a practical perspective, schools' management may learn about the affordances and constraints of using an LMS and use the information for decision making, strategy planning and possibly undertaking internal research to see how they are using their LMS. This research provides a basis of understanding a LMS in the context of South Africa specifically for primary schools.

## 6.  CONCLUSION

With the abrupt shift to remote teaching during ERT, the usage of a LMS in schools increased. As lockdown began, schools shut down and teaching moved online. An opportunity rose to study affordances and constraints faced when using a LMS for remote teaching, specifically for primary school learners ('Gen Z') – a largely ignored area of research as 'Gen Z' learners require new teaching methodologies to keep them engaged with content.

Several themes were generated for both affordances and constraints. Affordances were classified in relation to schools, learners, and teachers. Noteworthy affordances included using LMSs as notice boards and satisfying parents – the key stakeholders. Constraints were classified into financial, organizational, learner, teachers and technological. The digital divide in South Africa was a recurring theme while device and data costs were consistently a challenge.

From a practical perspective, this research will be useful to any primary school looking at introducing a LMS as part of their teaching. Existing users of a LMS will be able to learn from the experiences of other schools. For example, management of primary schools will understand the constraints their students may or already face and try to influence school policy or create mitigation strategies to counter constraints.

The research has limitations, the main one being that it is from a teachers' perspective only. There are several stakeholders in a school's setting that haven't contributed to this research such as students, parents, government and NGOs. Secondary sources while analysed were not presented here. Most of the schools used one brand of LMS thus adding an element of bias to the answers.

Some of the affordances and constraints raised are difficult to generalize as the researcher was unable understand if they were due to the specific brand of LMS.

To help understand more and bring more light on 'Gen Z' learners and LMSs, future work needs to be conducted from perspectives of other stakeholders involved in a school setting to understand the affordances and constraints faced. Lastly, researchers must include schools that use different LMSs to be able to generalize across technologies.